\documentclass{aipproc}
\layoutstyle{6x9}
\usepackage{epsfig}
\begin{document}

\title{Gamma-ray signatures of classical novae}

\author{Margarita Hernanz}
  {address={Institute for Space Studies of Catalonia (IEEC) and Instituto 
   de Ciencias del Espacio (CSIC), Edifici Nexus, C/Gran Capit\`a, 2-4, 
   E-08034 Barcelona, Spain},
   email={hernanz@ieec.fcr.es}}

\author{Jordi G\'omez-Gomar}
   {address={Institute for Space Studies of Catalonia (IEEC)},
    email={jgomez@ieec.fcr.es}}  

\author{Jordi Jos\'e}
  {address={Institute for Space Studies of Catalonia (IEEC) and Departament 
   de F\'{\i}sica i Enginyeria Nuclear (UPC), Avda. V\'{\i}ctor Balaguer, 
   s/n, E-08800 Vilanova i la Geltr\'u (Barcelona), Spain},
   email={jjose@ieec.fcr.es}}

\begin{abstract}
The role of classical novae as potential gamma-ray emitters is reviewed, 
on the basis of theoretical models of the gamma-ray emission from different 
nova types. The interpretation of the up to now negative results 
of the gamma-ray observations of novae, as well as the prospects for
detectability with future instruments (specially onboard INTEGRAL) are
also discussed.
\end{abstract}

\date{today}

\maketitle

\section{Introduction}
Classical novae are explosive phenomena occurring in close binary systems of 
the cataclysmic variable type. In these binaries, a normal main sequence 
star overflows its Roche lobe, transferring H-rich matter to its companion 
white dwarf 
star through an accretion disk. Matter accumulates on top of the degenerate 
white dwarf star, where it is gradually compressed and heated, until hydrogen 
reaches conditions for ignition. This ignition 
happens in a degenerate regime, thus leading to a thermonuclear runaway, 
because of the inability of matter to thermally readjust itself through 
expansion. During explosive hydrogen burning, radioactive nuclei (with 
lifetimes ranging from $\sim$100 s to $\sim 10^6$ s) are synthesized. The  
radioactive isotopes with lifetimes around 100 s, like $^{14}$O 
($\tau$=102 s), $^{15}$O ($\tau$=176 s) and $^{17}$F ($\tau$=93 s), are 
responsible for the explosion itself, because they can be transported by 
convection to the outer envelope, during the thermonuclear runaway (since 
$\tau_{\rm conv}<\tau$). These nuclei are prevented from destruction in the 
outer cooler shells, and their subsequent decay releases energy which is 
largely responsible for the expansion and large increase in luminosity of 
the nova.

Other radioactive isotopes synthesized in novae, with longer lifetimes, are 
responsible for the gamma-ray emission of these objects. Two types of emission 
are expected: prompt emission, related with e$^-$-e$^+$ annihilation (with 
e$^+$ coming from the decay of the short-lived $^{13}$N, $\tau$=862 s, and  
$^{18}$F, $\tau$=158 min) and long-lasting emission, caused by the decay of 
$^{7}$Be ($\tau$=77 days) and $^{22}$Na ($\tau$=3.75 yr). The prompt emission 
appears very early (before optical maximum, i.e., usually before nova 
discovery), has short duration (a couple of days) and consists of a 511 keV 
line plus a continuum below it (see 
below for details). The long-lasting emission consists of lines (either 478 
keV from $^{7}$Be decay or 1275 keV from $^{22}$Na decay), lasting around 2 
months and 3 years, respectively.

The potential role of classical novae as sources of gamma-ray emission was 
pointed out long ago \cite{CH74, Cla81, LC87},
but detailed models combining both the explosion modeling and 
the production and propagation of gamma-rays are more recent \cite{Her97a, 
Her97b, Gom98, Her99}.
Up to now, there have been unsuccessful attempts to detect gamma-ray 
emission from novae. Efforts have 
been made mainly to detect the $^{22}$Na line, at 1275 keV, with the COMPTEL 
instrument onboard the Compton Gamma-Ray Observatory, CGRO \cite{Iyu95, Iyu99}.
Previous attempts to detect 
the $^{7}$Be line, at 478 keV, and the 1275 keV line were made with the 
GRS instrument 
onboard the Solar Maximum Mission, SMM, satellite \cite{Har91}
All these efforts 
have only provided upper limits, fully compatible with our theoretical 
predictions \cite{JH98, JCH99}

Other attempts have concentrated on the annihilation emission (511 keV line 
plus continuum below it), with large field of view instruments, like
WIND/TGRS \cite{Har99} and CGRO/BATSE \cite{Her00}, 
without success and, again, with upper limits compatible with 
theoretical predictions. The possible detection of this type of emission 
from novae with the CGRO/BATSE instrument had been pointed out by Fishman et 
al. (1991) prior to CGRO launch. The sensitivities of the instruments 
were too low to detect the emission, which is more intense than that in the 
478 and 1275 keV lines but has much shorter duration. 
In addition to search for gamma-ray emission in particular 
objects, there have been attempts to look for the Galactic accumulated 
emission at 478 and 1275 keV, both with CGRO/OSSE and SMM/GRS 
\cite{Lei88, Har91, Har96}. In this case, more flux is accumulated 
since more sources are contributing, because the typical period between two 
succesive nova explosions in the Galaxy is shorter than the lifetimes of 
$^{7}$Be and $^{22}$Na. But again not enough sensitivity was available. We 
have recently made predictions about the detectability of this accumulated 
emission with INTEGRAL/SPI \cite{Jea00}; the cumulative emission around the 
Galactic center has some chance of being detected with SPI, during the 
deep survey of the central radian of the Galaxy (or, at least, better upper limits 
than those of SMM or COMPTEL are expected).

\section{Gamma-ray emission: lines and continuum}

A summary of the main radioactive nuclei synthesized in novae is shown in 
table \ref{radioac1}. It is important to stress that these nuclei are not 
produced in the same amounts in all the nova types, since their synthesis is 
closely related to the nuclear paths followed by the nova during its 
evolution. These paths depend on the initial chemical composition of the 
accreted envelope, which is related to that of the underlying white dwarf 
core, because some mixing between the core and the envelope should be invoked 
in order to explain the observed abundances of novae. It turns out that 
CO novae are the 
main producers of $^{7}$Be, whereas ONe novae are responsible for $^{22}$Na 
synthesis. In table \ref{radioac2} we show some examples of nova models, 
with their relevant yields of radioactive isotopes. The specific kinetic 
energy of the ejecta is also shown for completeness. These results have been 
obtained by means of a hydrodynamic code, which computes the nova evolution 
from the accretion phase up to the explosion and ejection of the envelope (see 
Jos\'e \& Hernanz 1998 for details). The $^{18}$F yields still suffer from 
some uncertainty, mainly because of the 
not well known $^{18}$F(p,$\alpha$) reaction (see \cite{Coc00} for a recent 
analysis).

\begin{figure}
\setlength{\unitlength}{1cm}
\begin{picture}(18,8)
\put(1,-2){\makebox(9,10){\epsfxsize=8cm \epsfbox{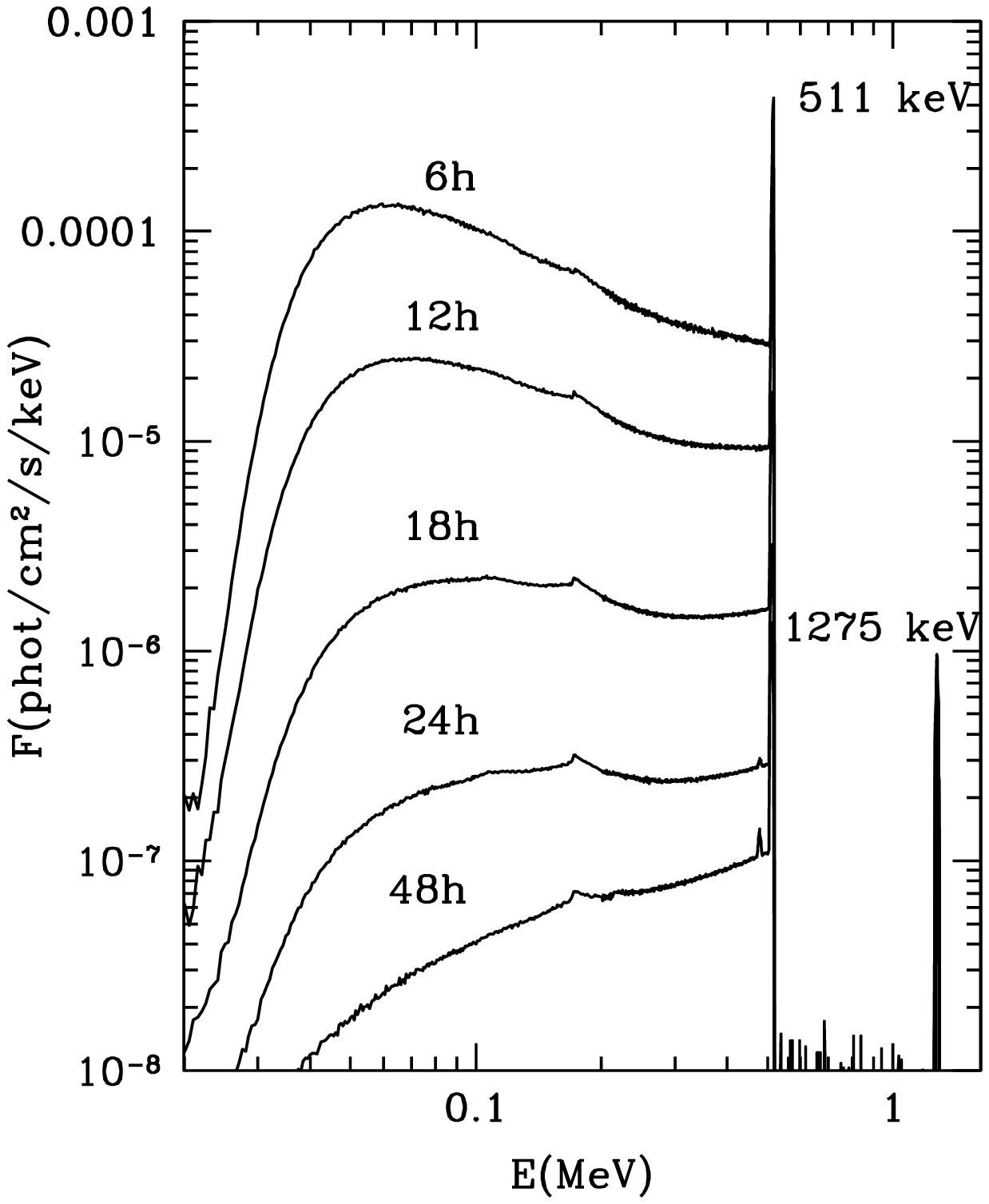}}}
\put(9,-2){\makebox(9,10){\epsfxsize=8cm \epsfbox{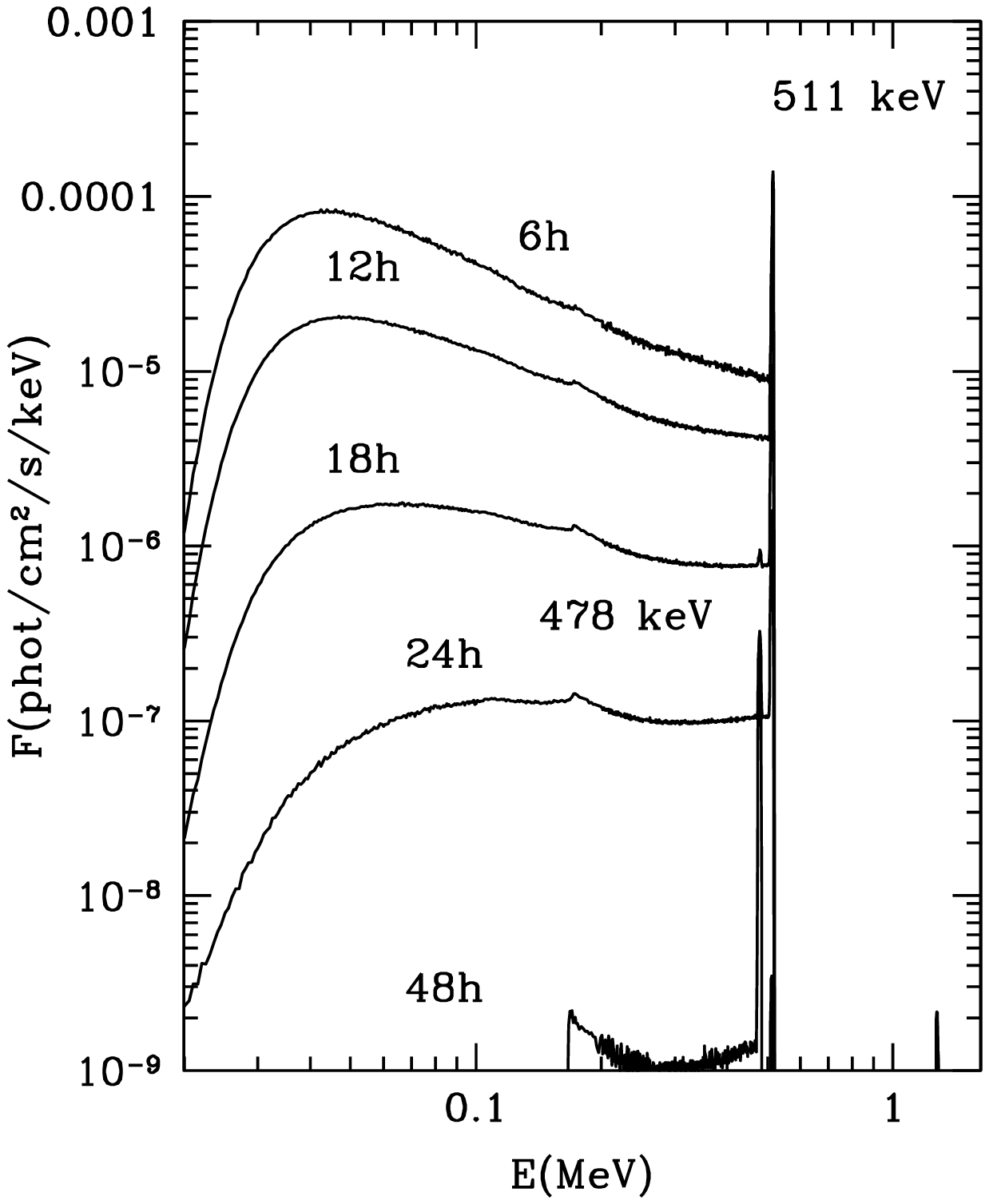}}}
\end{picture}
\caption{(Left) Gamma-ray spectra for an ONe nova of 1.25M$_\odot$, at 
different epochs after the outburst (defined as the peak temperature time) 
and at distance 1 kpc. (Right) Same for a CO nova of 1.15M$_\odot$}
\label{spectra}
\end{figure}

The gamma-ray output of a particular nova model at different epochs after the 
outburst (defined as the epoch of peak temperature), has been 
computed with a Monte Carlo code, which handles gamma-ray production and 
transfer in the expanding envelope (see \cite{Gom98} for 
details), with properties derived from the hydro code models. In figure 
\ref{spectra} we show the spectral evolution of a CO and an ONe nova 
(M$_{wd}$=1.15 and 1.25 M$_\odot$, respectively), at distance 1 kpc. For all 
models there is a continuum between (20-30) and 511 keV, and a line at 511 
keV ($\sim 8$ keV full width half-maximum, FWHM), with 
intensities decreasing very fast \cite{Her99}. The 511 
keV line comes from the direct annihilation of positrons and from the 
positronium (in singlet state) emission, whereas the continuum originates in 
both the positronium continuum (triplet state positronium) and the 
Comptonization of photons emitted in the 511 keV line. There is a cutoff of 
the continuum at low energies (20-30 keV, depending on the chemical 
composition), related to photoelectric absorption, which acts as a sink of the 
Comptonized photons. In addition to 
this prompt and short-duration emission, there is a longer duration gamma-ray 
output, consisting of a line at 478 keV ($\sim$(3-8) keV FWHM), in CO novae, 
or at 1275 keV ($\sim$20 keV FWHM), in ONe novae. The general trends for other 
CO and ONe models are similar to those shown here. It is worth noticing that 
models with lower masses are more opaque (i.e., the 0.8 M$_\odot $ CO nova), 
because of the smaller expansion velocities (see table \ref{radioac2}).
 
\begin{figure}
\setlength{\unitlength}{1cm}
\begin{picture}(18,8)
\put(1,-2){\makebox(9,10){\epsfxsize=8cm \epsfbox{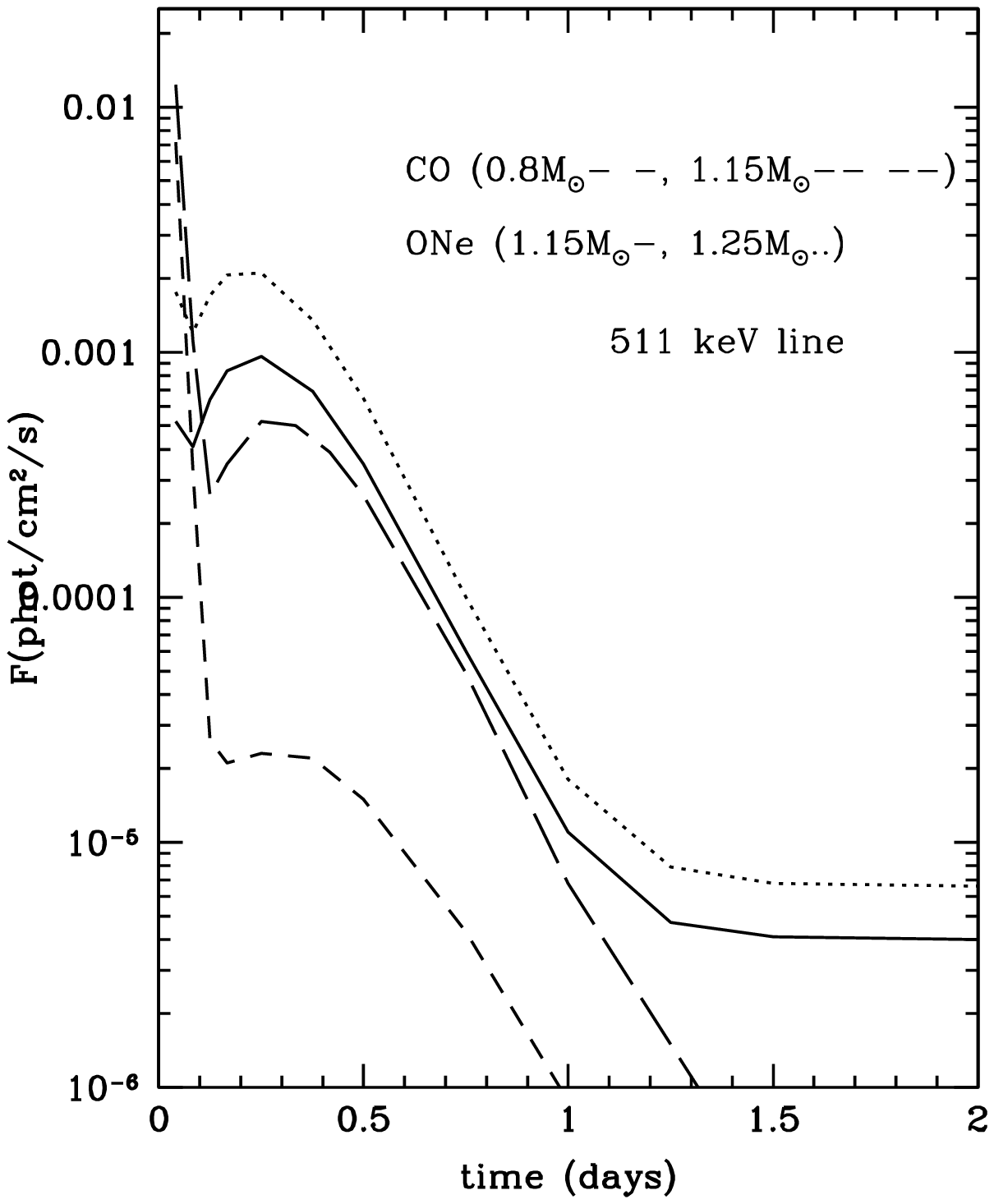}}}
\put(9,-2){\makebox(9,10){\epsfxsize=8cm \epsfbox{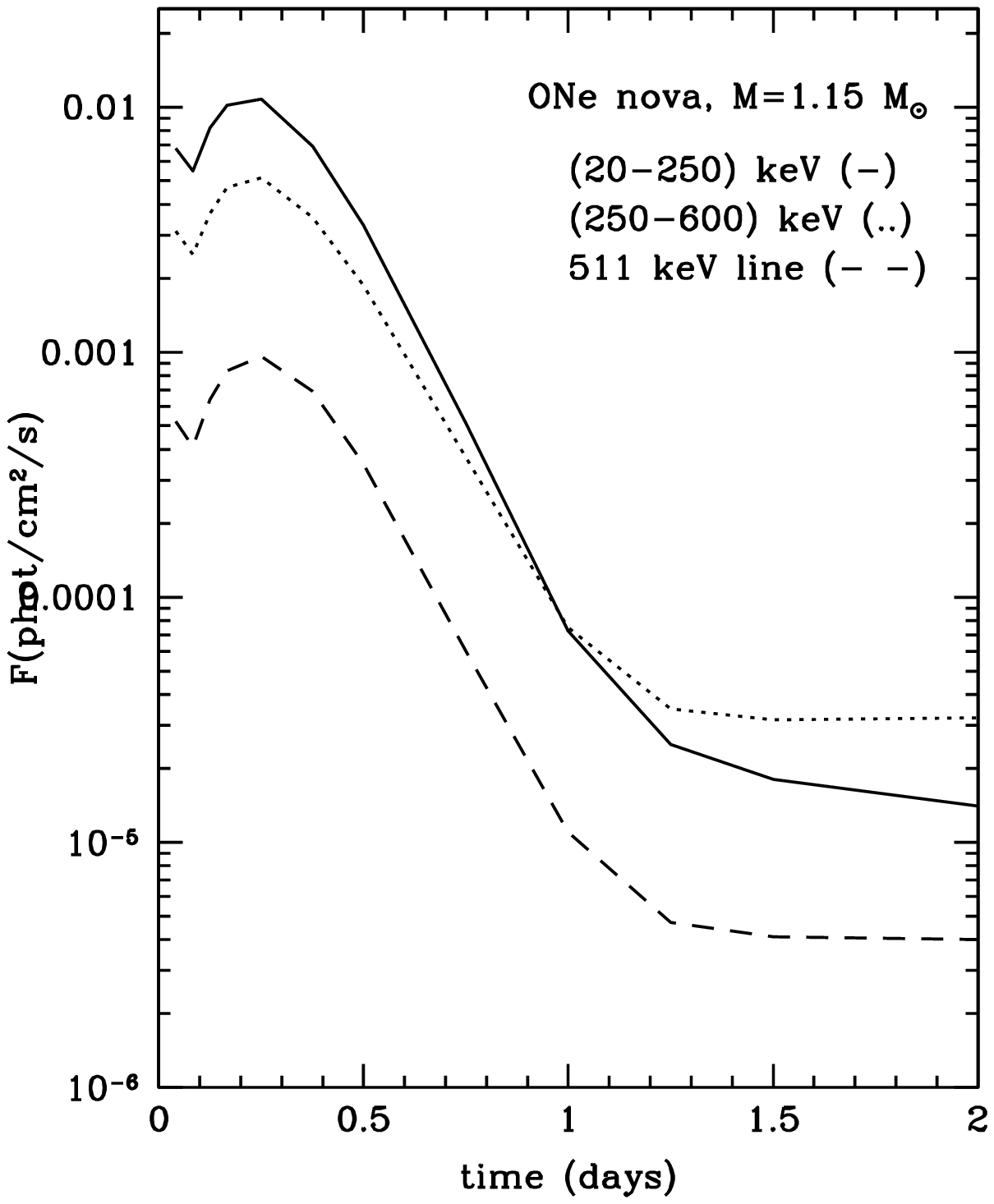}}}
\end{picture}
\caption{(Left) Light curves for the 511 keV line of the 4 nova models shown 
in table \ref{radioac2}, placed at a distance of 1 kpc . 
(Right) Continuum light curves for the ONe nova of 1.15 M$_\odot$ at the same 
distance.}
\label{lcannihil}
\end{figure}

\begin{figure}
\setlength{\unitlength}{1cm}
\begin{picture}(18,8)
\put(1,-2){\makebox(9,10){\epsfxsize=8cm \epsfbox{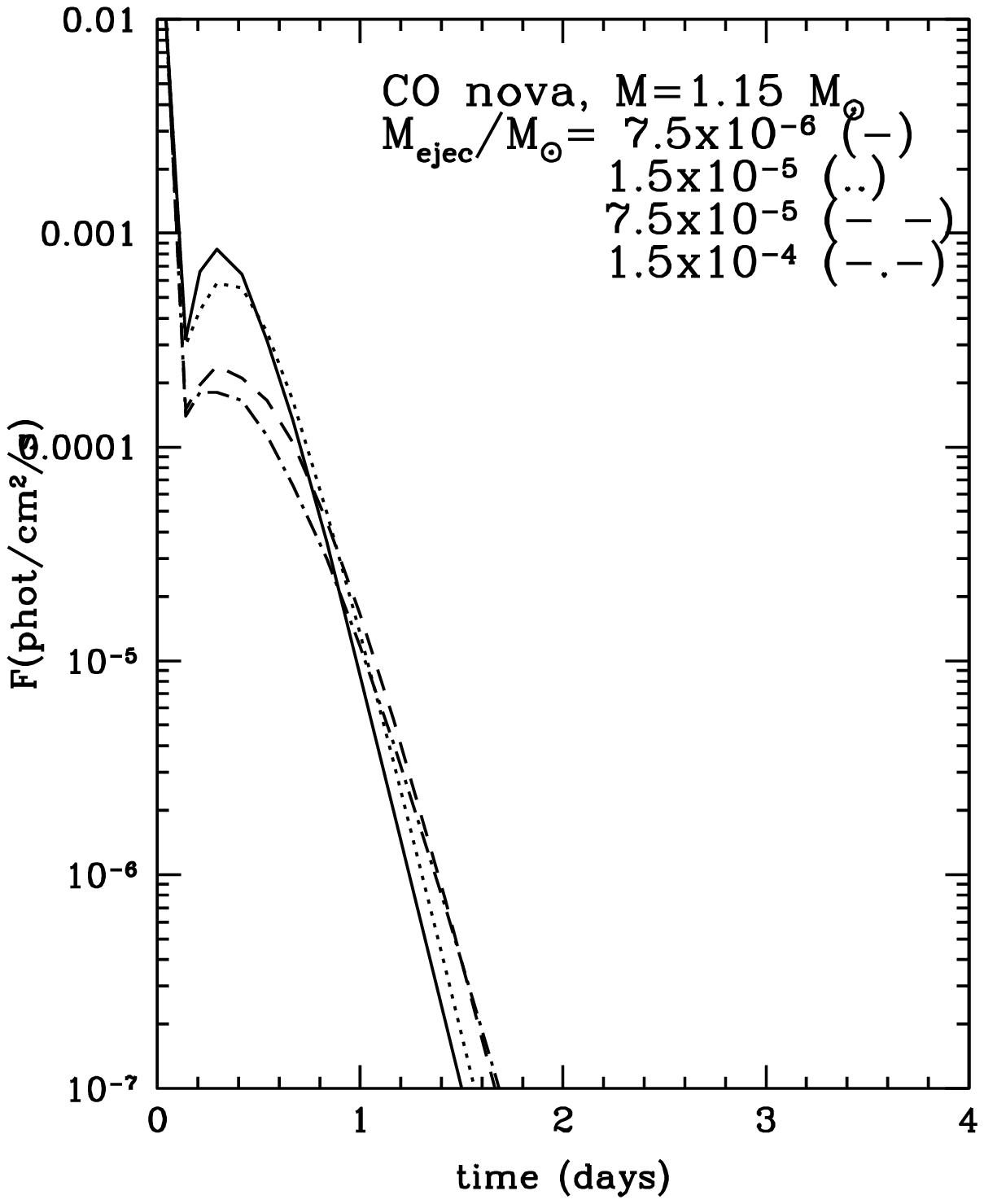}}}
\put(9,-2){\makebox(9,10){\epsfxsize=8cm \epsfbox{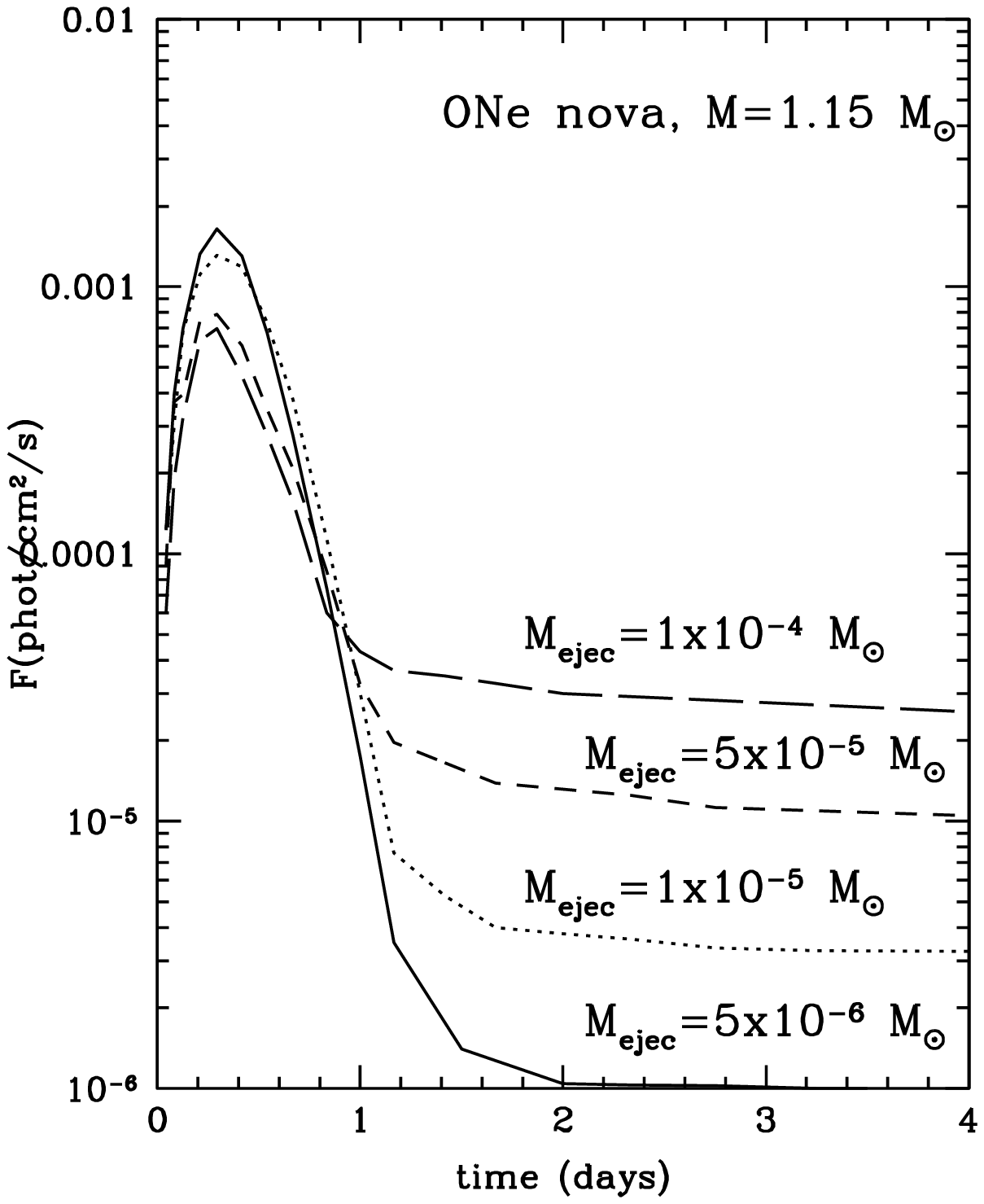}}}
\end{picture}
\caption{(Left) Light curves for the 511 keV line for a CO nova of 
1.15 M$_\odot$, for a range of ejected masses. (Right) Same for an ONe nova of 
1.15 M$_\odot$. Distance is 1 kpc. }
\label{paramm}
\end{figure}

\begin{figure}
\setlength{\unitlength}{1cm}
\begin{picture}(18,8)
\put(1,-2){\makebox(9,10){\epsfxsize=8cm \epsfbox{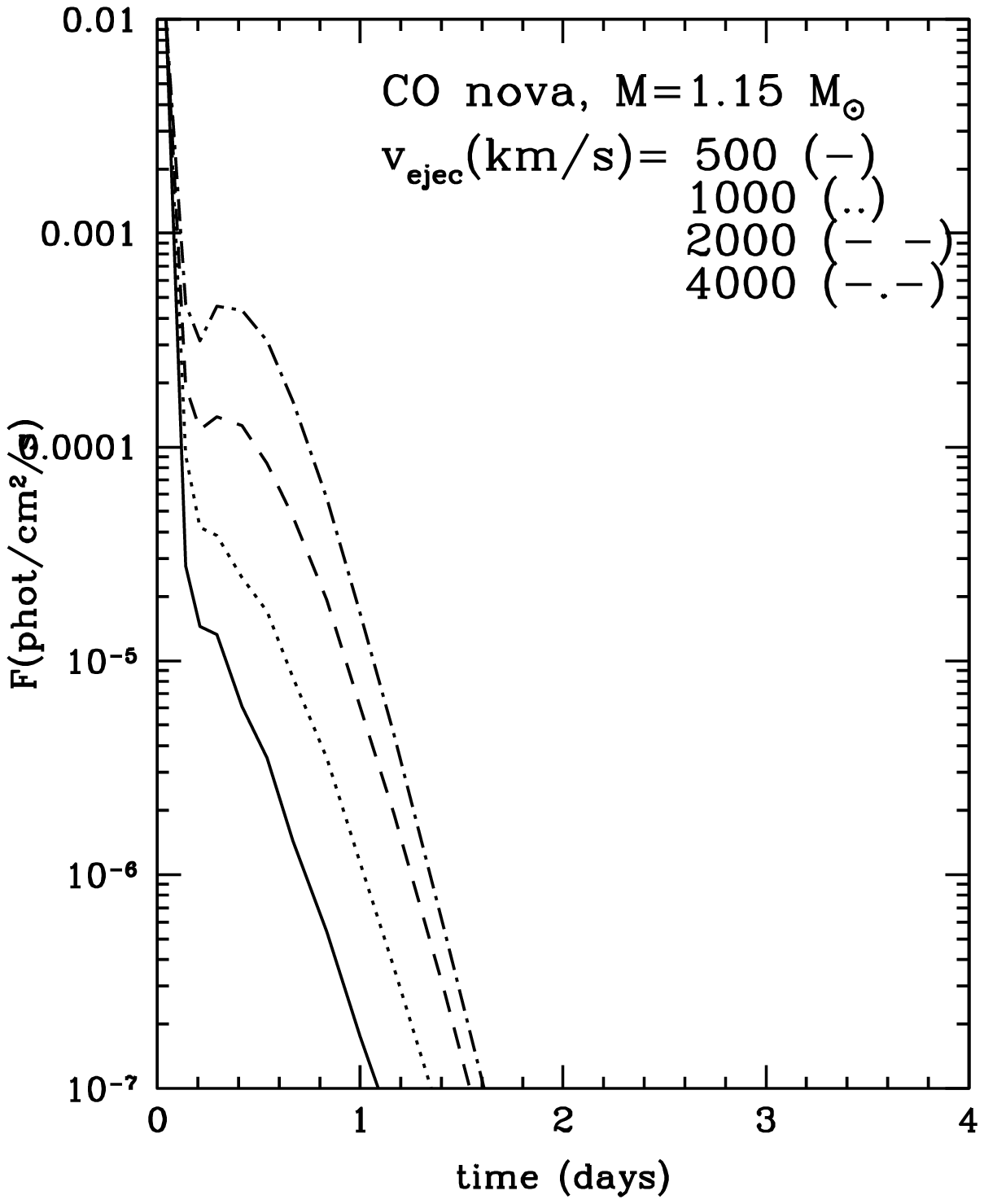}}}
\put(9,-2){\makebox(9,10){\epsfxsize=8cm \epsfbox{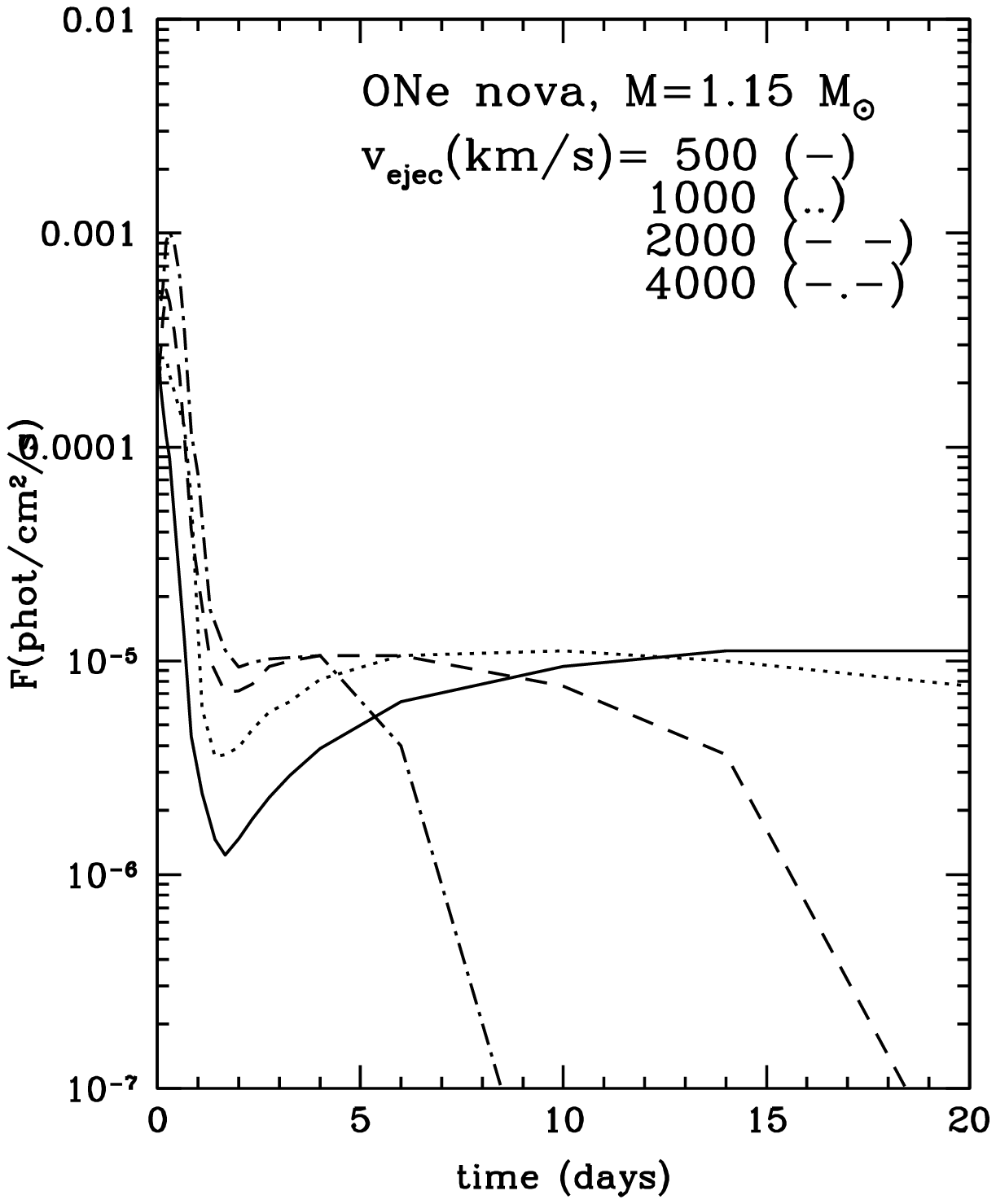}}}
\end{picture}
\caption{(Left) Light curves for the 511 keV line for a CO nova of 
1.15 M$_\odot$, for a range of parametrized velocities of the ejecta. The value 
indicated corresponds to the outermost shell. 
(Right) Same for an ONe nova of 1.15 M$_\odot$. Distance is 1 kpc.}
\label{paramv}
\end{figure}

\begin{figure}
\setlength{\unitlength}{1cm}
\begin{picture}(18,8)
\put(1,-2){\makebox(9,10){\epsfxsize=8cm \epsfbox{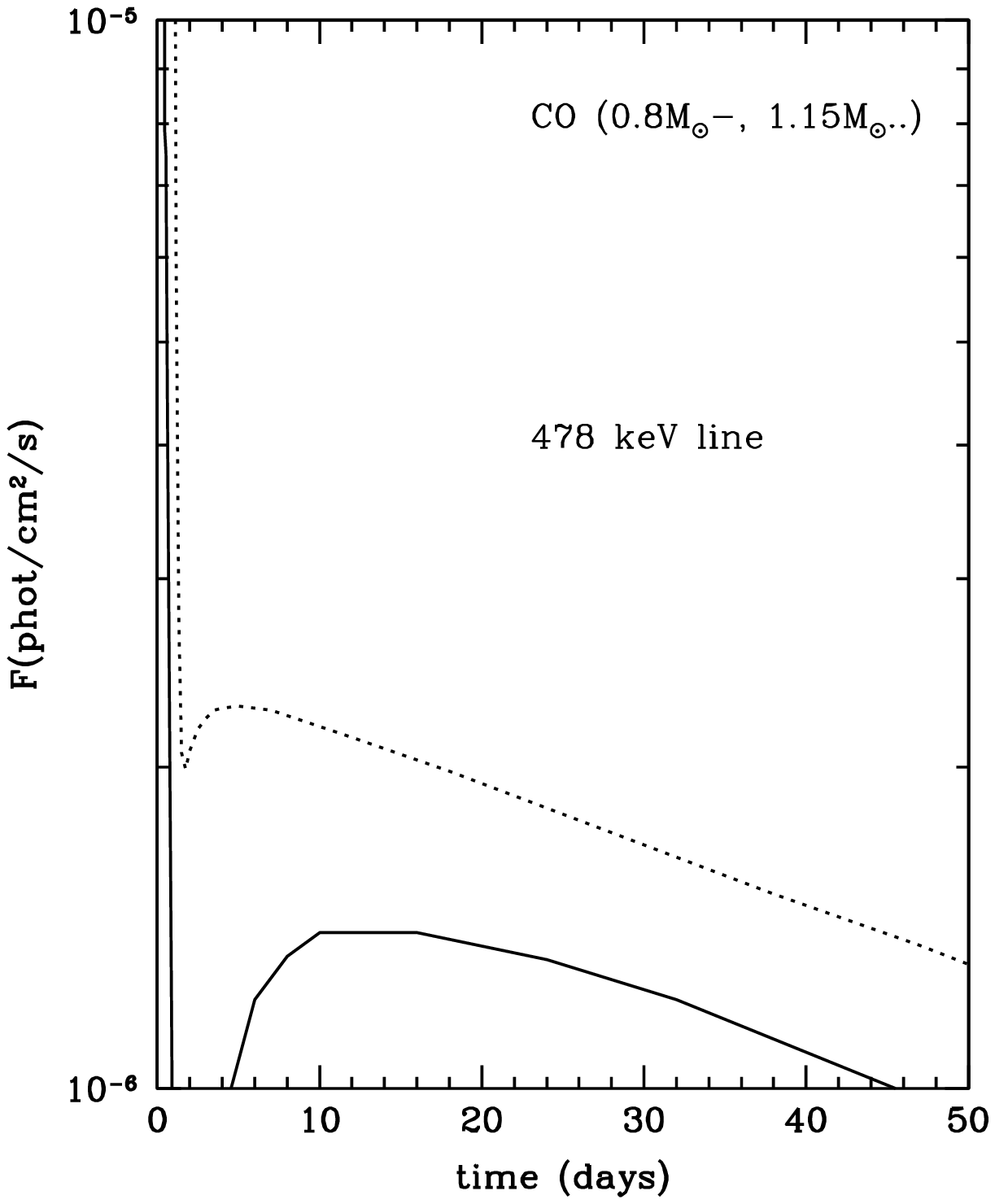}}}
\put(9,-2){\makebox(9,10){\epsfxsize=8cm \epsfbox{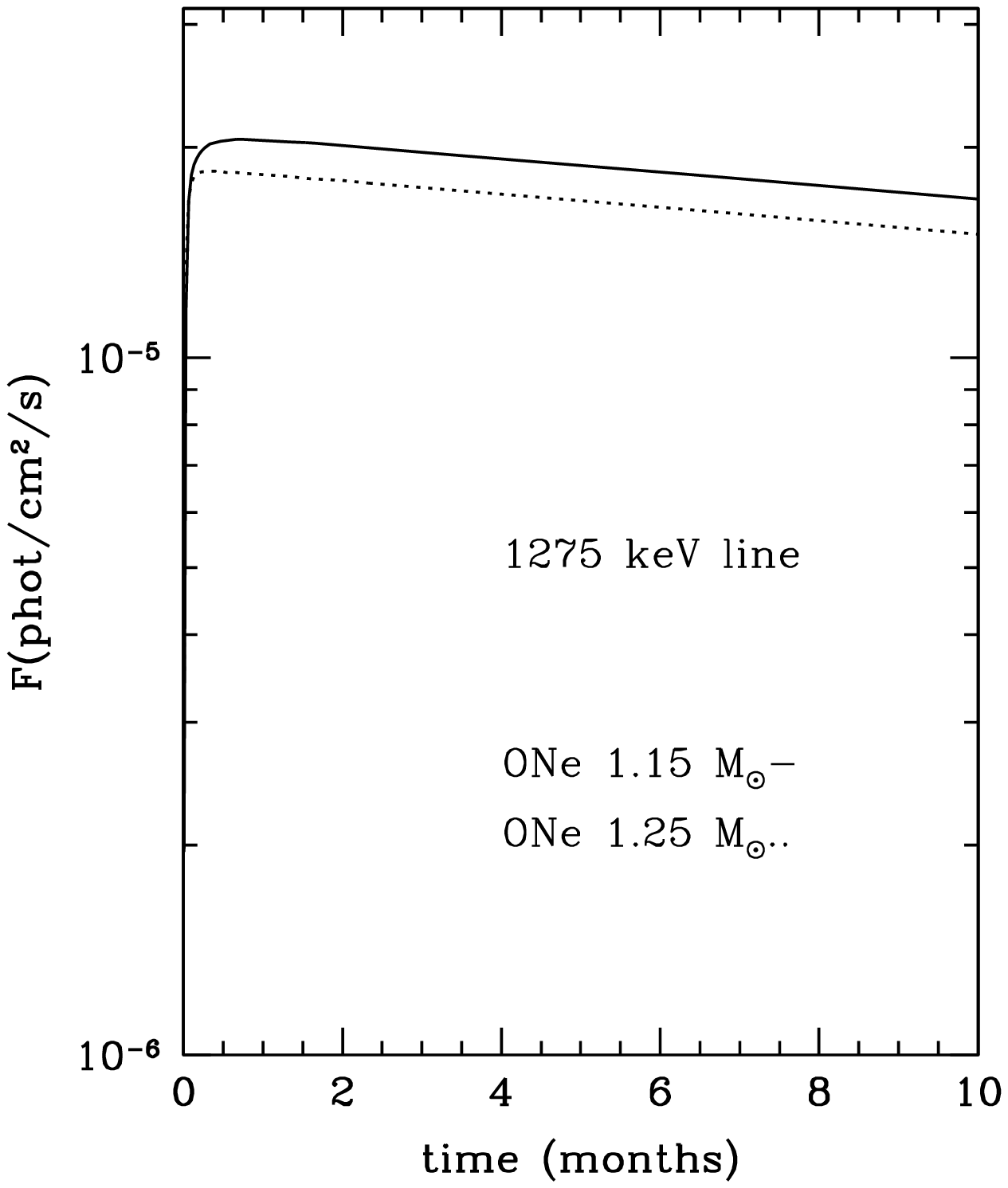}}}
\end{picture}
\caption{(Left) Light curves for the $^{7}$Be line (478 keV) for two CO nova 
models. (Right) Light curves for the $^{22}$Na (1275 keV) for two ONe models. 
Distance is 1 kpc.}
\label{lclines}
\end{figure}

The light curves for the different types of emission are shown in figures 
\ref{lcannihil}, \ref{paramm}, \ref{paramv} and \ref{lclines}. 
Figure \ref{lcannihil} shows the
light curve of the 511 keV line (FWHM between 3 and 8 keV) for all models, 
and those of different energy bands in the continuum, for an ONe nova. The 
continuum emission at energies lower than 511 keV dominates, being the band 
between 20 
and 250 keV the one with the highest flux (but also the one which decreases 
faster, as can also be seen in figure \ref{spectra}). This prompt emission 
gives a direct insight of the dynamics of the expanding envelope, as well as 
information about its content on the radioactive nuclei $^{13}$N and $^{18}$F. 
In the case of ONe novae, there is also the contribution of positrons from 
$^{22}$Na decay, which produces smaller fluxes but lasts a longer time (up 
to complete transparency of the envelope, which occurs at around 1 week 
after peak temperature, the exact value depending on the expansion velocities of 
the envelope). 

We have analyzed the influence of the mass and the velocity of the ejecta on 
the prompt emission, by means of some extra models, in which we scale 
either the mass of the ejecta or its terminal velocity. These 
are in some way not self-consistent models, because they are not the result of 
evolutionary calculations, but they are good for illustrative purposes. Figure 
\ref{paramm} shows the 511 keV line light curves for a CO and an ONe nova 
(both of 1.15 M$_\odot$), for a range of parametrized ejected masses (the 
value obtained in the evolutionary model is shown in table \ref{radioac2}). The 
effect of increasing the ejected mass is twofold, depending on the epoch. 
At early times, the 
larger the mass the lower the flux, because of the increasing opacity. On the 
contrary, later on the opacity doesn't play an important role, and the larger 
the mass the larger the flux, because of the larger amount of radioactive 
isotopes. It is worth reminding that in ONe novae the emission 
lasts longer than in CO ones (see figure \ref{paramm}, right), because of the 
contribution of the e$^+$ from $^{22}$Na decay. 

The influence of the velocity of the ejecta is shown in figure \ref{paramv}. 
At early times, 
larger velocities imply larger transparency and thus larger fluxes (both for 
CO and ONe novae). At later times (after $\sim$ 2days), only the case 
of ONe novae is relevant, since there are still e$^+$ from $^{22}$Na decay; 
then, the larger the velocity the earlier the flux disappears, because the 
envelope becomes transparent before, thus allowing  e$^+$ to freely escape 
(see figure \ref{paramv}, right). 
This facts demonstrates again that the analysis of the prompt gamma-ray 
emission of classical novae would provide a great deal of information about 
the dynamics of the expanding envelope, as well as about the ratio between 
its $^{18}$F and $^{22}$Na contents.

In figure \ref{lclines} we display the light curves of the 478 keV line, for 
the two CO novae from table \ref{radioac2}, and those of the 1275 keV line, 
for the ONe novae in table \ref{radioac2}. These light curves show a first 
phase of increasing flux, related to the increasing transparency of the 
envelope, followed by the characteristic exponential decay phase, when the 
envelope is already transparent. The light curve of the 478 keV line shows in 
addition an intense peak at early times, which comes from the Comptonization 
of the 511 keV photons (see above). The fluxes of the 478 and 1275 keV lines 
during the exponential decay phase, directly reflect the amount of $^{22}$Na 
and $^{7}$Be in the envelope. 

\section{Prospects for detectability of individual novae}

In order to predict detectability distances of the gamma-ray emission from 
novae, the abovementioned light curves for the different types of emission 
have been used. The fluxes are 
quite small, leading to detectability distances with INTEGRAL/SPI of around 
1 kpc, for the 1275 keV line, and 0.5 kpc, for the 478 keV line, for the 
nominal observation time of $10^6$s (see table 
\ref{SPIdetect} for exact values). Concerning the 511 keV line and the 
continuum, detectability distances with SPI are around 3 kpc (see 
table \ref{SPIdetect}), adopting 10 h of observation time, 
starting 5 h after peak temperature. For the continuum we have adopted the 
range (170-470) keV, which is optimal for SPI, since it avoids the 478 keV 
line and the low energies, where the background is too high. The width of 
the lines has been fully taken into account to derive all the detectability 
distances. As it is known, the instrument INTEGRAL/SPI will have a very good 
spectral resolution, which means that its nominal sensitivity for narrow 
lines is degraded when they are broad. 

Our time origin in the figures is at peak temperature, which happens 
before the maximum in visual luminosity. The time interval between 
peak temperature and maximum in visual luminosity depends on the particular 
nova model, mainly on its speed class (rate of decline of the visual 
luminosity). It ranges from some days to some weeks, but its exact value 
is difficult to establish, because novae are usually discovered at or after  
visual maximum. Therefore, the epoch of peak temperature is close to peak 
gamma-ray luminosity (corresponding to the e$^-$-e$^+$ annihilation 
emission), but it is not reachable from visual observations. 
The early appearence, before optical detection, of the prompt gamma-ray 
emission from novae, makes its detection with SPI problematic. It will be only 
possible if a close enough nova falls in the field of view of the instrument 
when it is 
doing another observation (i.e., during the Galactic plane survey -GPS- or 
during the Galactic center deep exposure -GCDE). We have also considered 
alternative ways to detect this intense emission, by means of the SPI shield, 
which provides a large detection area with a wide field of view, but without 
spectroscopic capability \cite{Jea99}.
In summary, the prompt gamma-ray e$^-$-e$^+$ annihilation  
emission can almost only be detected with wide field of view intruments 
scanning all the sky very often (like the future EXIST, MEGA, Advanced Compton 
Telescope). Up to now, ``a posteriori'' analyses (provided that 
there was some observation of the right field at the right moment) of the 
CGRO/BATSE \cite{Her00} and WIND/TGRS \cite{Har99} data 
have been performed; the negative results are fully compatible with our 
theoretical predictions and are related to the not enough sensitivity of these 
instruments. 

\section{Discussion}
 
The main factor affecting detectability of novae is distance (see table 
\ref{SPIdetect}), but the distances of novae are not easy to determine 
accurately. 
The visual luminosity (i.e., the absolute visual magnitude) of a classical 
nova at maximum is not directly correlated with its amount of the 
radioactive nucleus $^{22}$Na, or any other radioactive nucleus (in contrast 
with SNIa, where $^{56}$Ni is responsible for both the visual and the gamma-ray 
luminosities at early times). Therefore, some other characteristic, such as 
apparent visual 
magnitude, should be used as distance indicator. But, as 
with any cosmic object, novae which are apparently bright visually can be 
farther away than novae which are dim, if the visual extinction (intrinsic 
plus interstellar) of the apparently bright object is much smaller than that 
of the apparently dim object. Once the preliminary visual light curve and 
visual extinction are obtained, a distance determination is possible through 
indirect methods, which suffer from large uncertainties. They depend on 
various not well known nova properties. First, the
empirical relationship between absolute magnitude at maximum, $M_V^{max}$, and 
speed class of the nova (MMRD relation); the speed class is measured by the 
time of decline of the visual magnitude by 2 or 3 magnitudes ($t_2$ or $t_3$). 
Second, the visual extinction of the nova, $A_V$, which has intrinsic plus 
interstellar contributions; the latter varies a lot depending on the 
location of the nova in the Galaxy.

\begin{figure}
\setlength{\unitlength}{1cm}
\begin{picture}(18,8)
\put(2,-2){\makebox(9,10){\epsfxsize=8cm \epsfbox{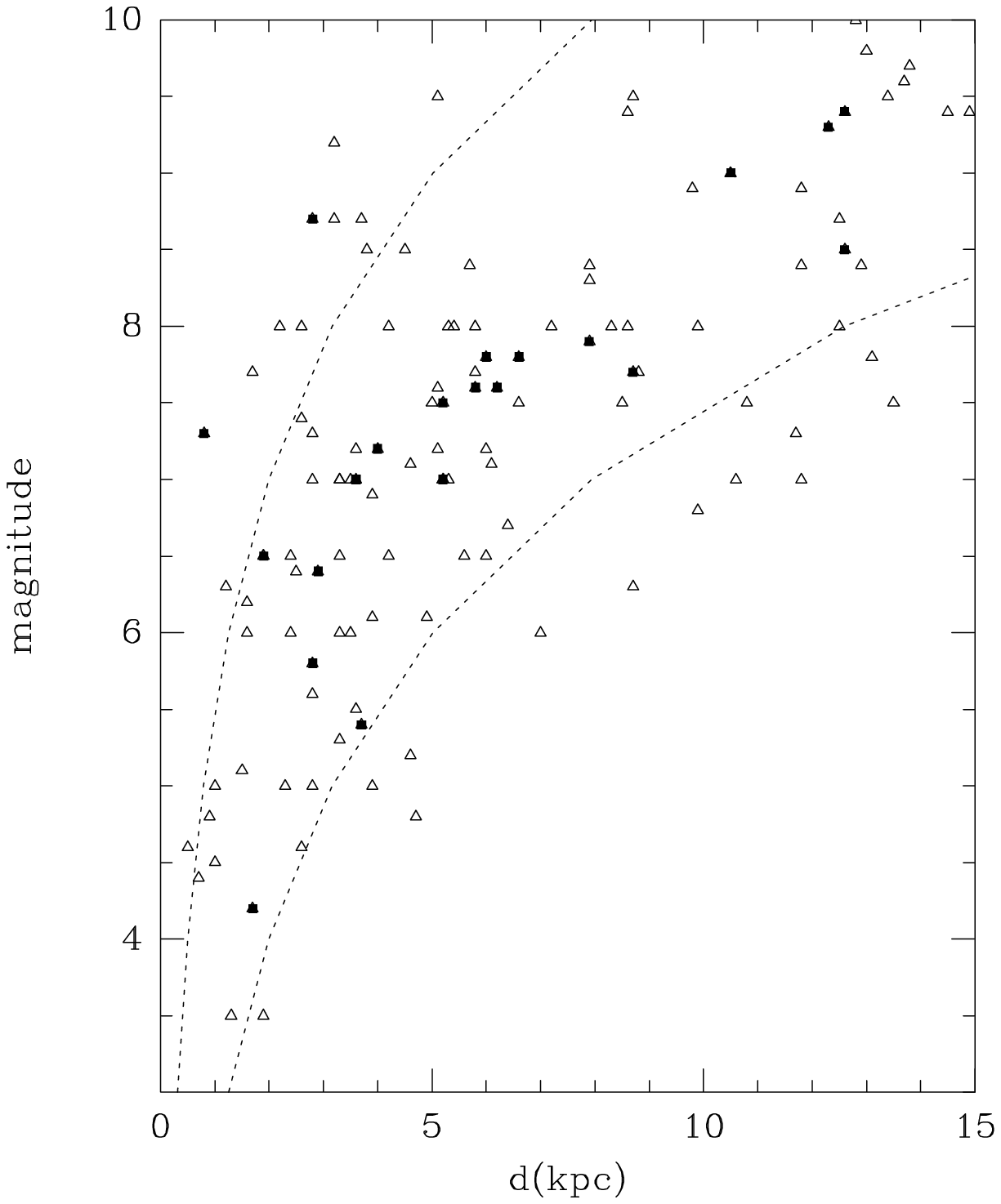}}}
\put(10,-2){\makebox(9,10){\epsfxsize=8cm \epsfbox{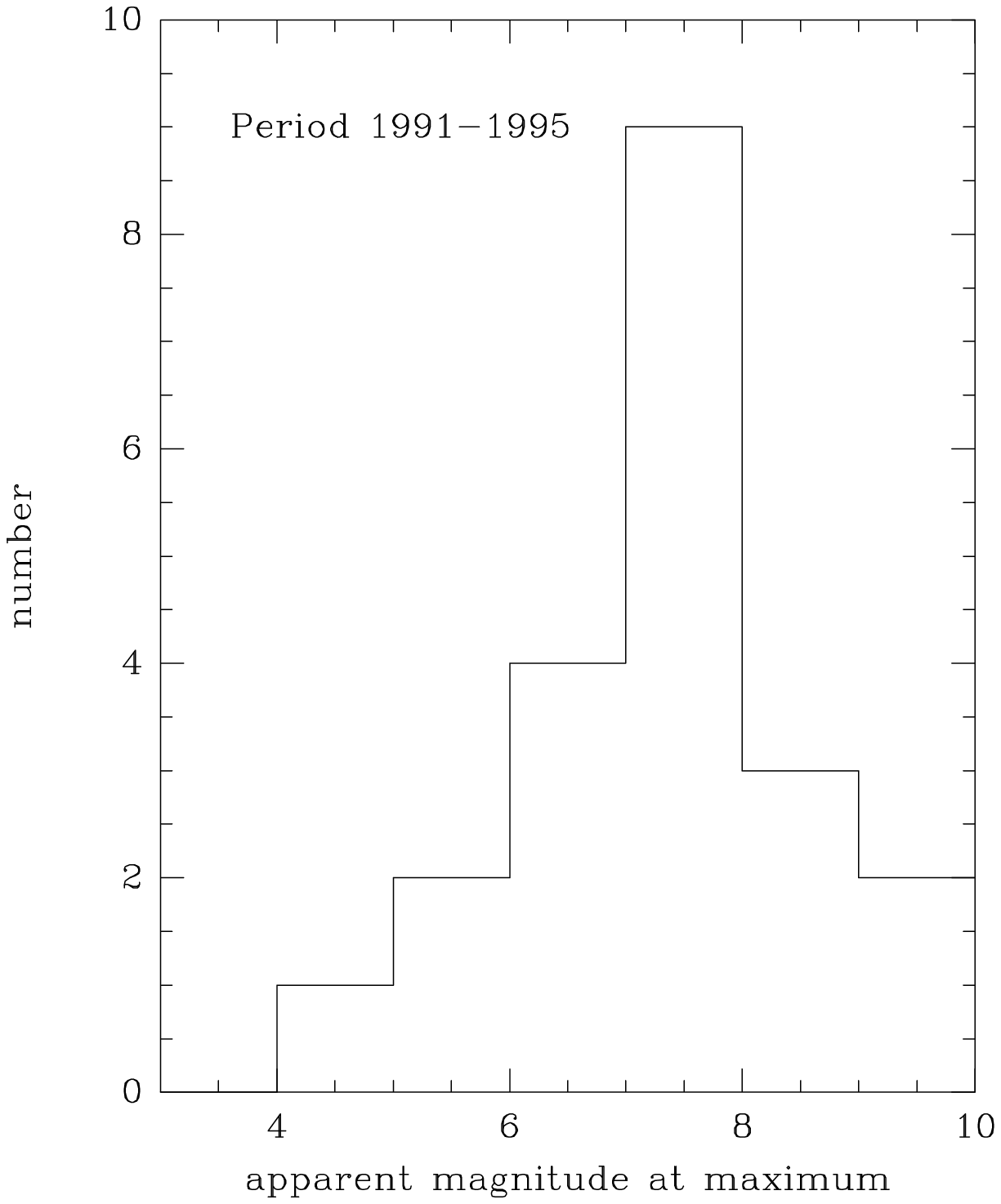}}}
\end{picture}
\caption{(Left) Apparent visual magnitudes at maximum, 
$m_V^{max}$, versus distances. Filled squares correspond to the 1991-1995 
period and open 
triangles to the 1901-1990 period. The dashed curves represent the
$m_V^{max}$ vs. distance relationship obtained for an absolute 
$M_V^{max}=-7.5$ (typical for novae) and a range of visual extinctions (from 
right to left $A_V=0$ and $A_V=3$ magnitudes). (Right) Histogram of novae 
apparent magnitudes at maximum, for the novae in the period 1991-1995.}
\label{magnitudes}
\end{figure}

Once $M_V^{max}$ and $A_V$ are known, the derivation of the distance from the 
apparent magnitude at maximum, $m_V^{max}$, is straightforward. Therefore, 
the main uncertainties affecting distance determinations are: general 
validity of the empirical $M_V^{max}$-$t_2$ (or $t_3$) relationship, 
determination of $A_V$, in addition to the determination of $t_2$ (or $t_3$) 
and of $m_V^{max}$ (often it is not known if the nova has been caught at the 
maximum or after it) from the observations. 
In figure \ref{magnitudes}, we show a $m_V^{max}$-distance diagram, for novae 
discovered in the last century (up to 1995). The data shown are 
taken from the samples of Shafter \cite{Sha97}. We have superimposed two curves 
indicating the 
apparent magnitudes at maximum, $m_V^{max}$, one could expect, provided that 
novae are standard candles with absolute magnitude at maximum 
$M_V^{max}$=-7.5, and that visual extinction, $A_V$, ranges from 0 to 3 
magnitudes. For distances up to 1 kpc, $m_V^{max}$ should be smaller 
(brighter) than 5.5 
(for 3 kpc, $m_V^{max}$ ranges from 8 to 5, or brighter if  $M_V^{max}$ is 
$<-7.5$). If we include novae after 1995, two outstanding points at 
$m_V^{max}$=2.8 and 4, and d$\sim$ 2 and 4 kpc (Nova Vel 1999 and Nova Aql 
1999b, respectively) would appear (with $M_V^{max} < -7.5$; Nova 
Vel 1999 probably had $M_V^{max} \sim -8.7$ (IAUC 7193)), 
in addition to more ``normal'' points with distances larger than 5 kpc and 
$m_V^{max}$ larger than 8. The number of novae discovered  
during the period 1991-1995 versus $m_V^{max}$ is also shown in figure 
\ref{magnitudes}.

\begin{figure}
\setlength{\unitlength}{1cm}
\begin{picture}(18,8)
\put(2,-2){\makebox(9,10){\epsfxsize=8cm \epsfbox{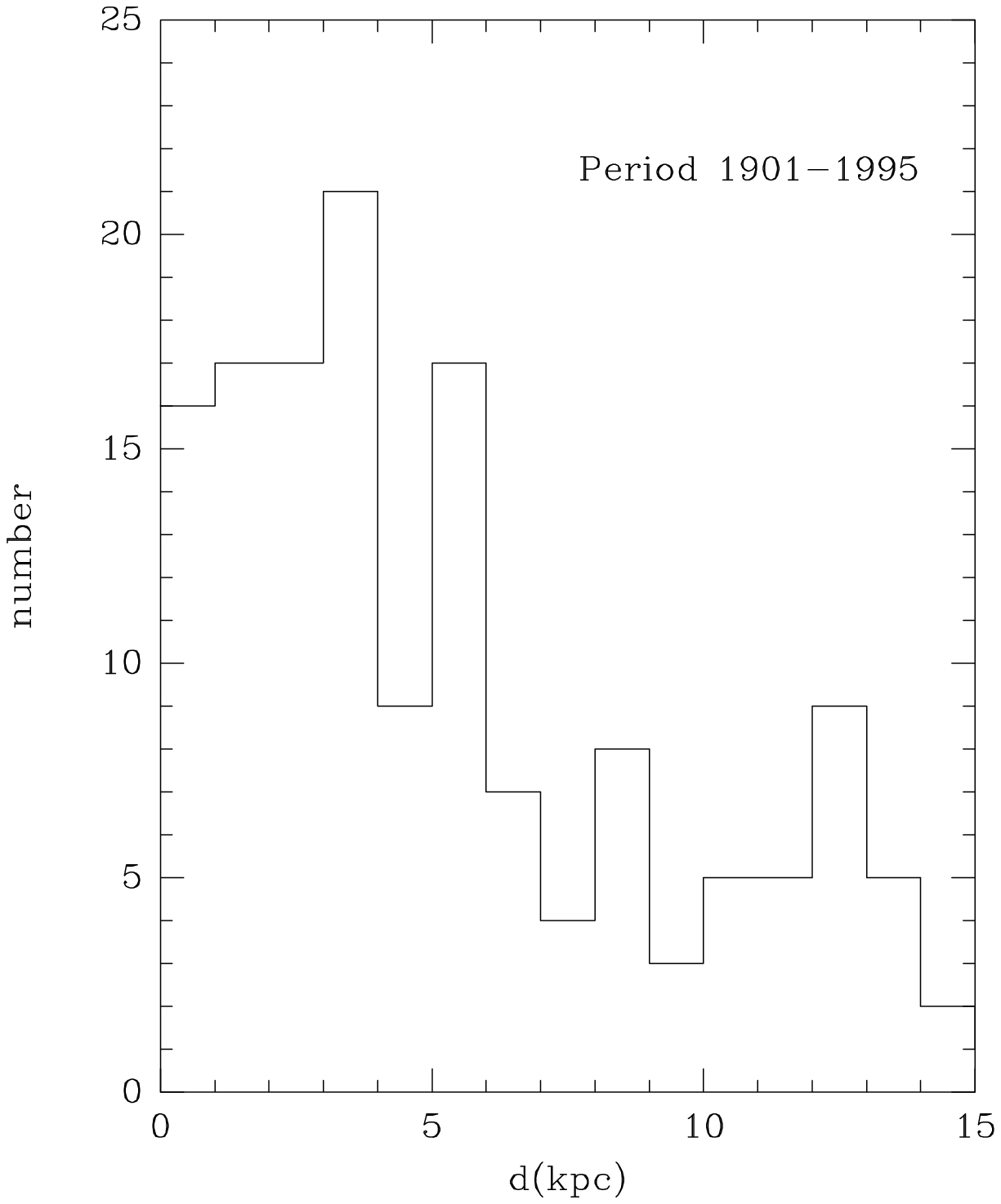}}}
\put(10,-2){\makebox(9,10){\epsfxsize=8cm \epsfbox{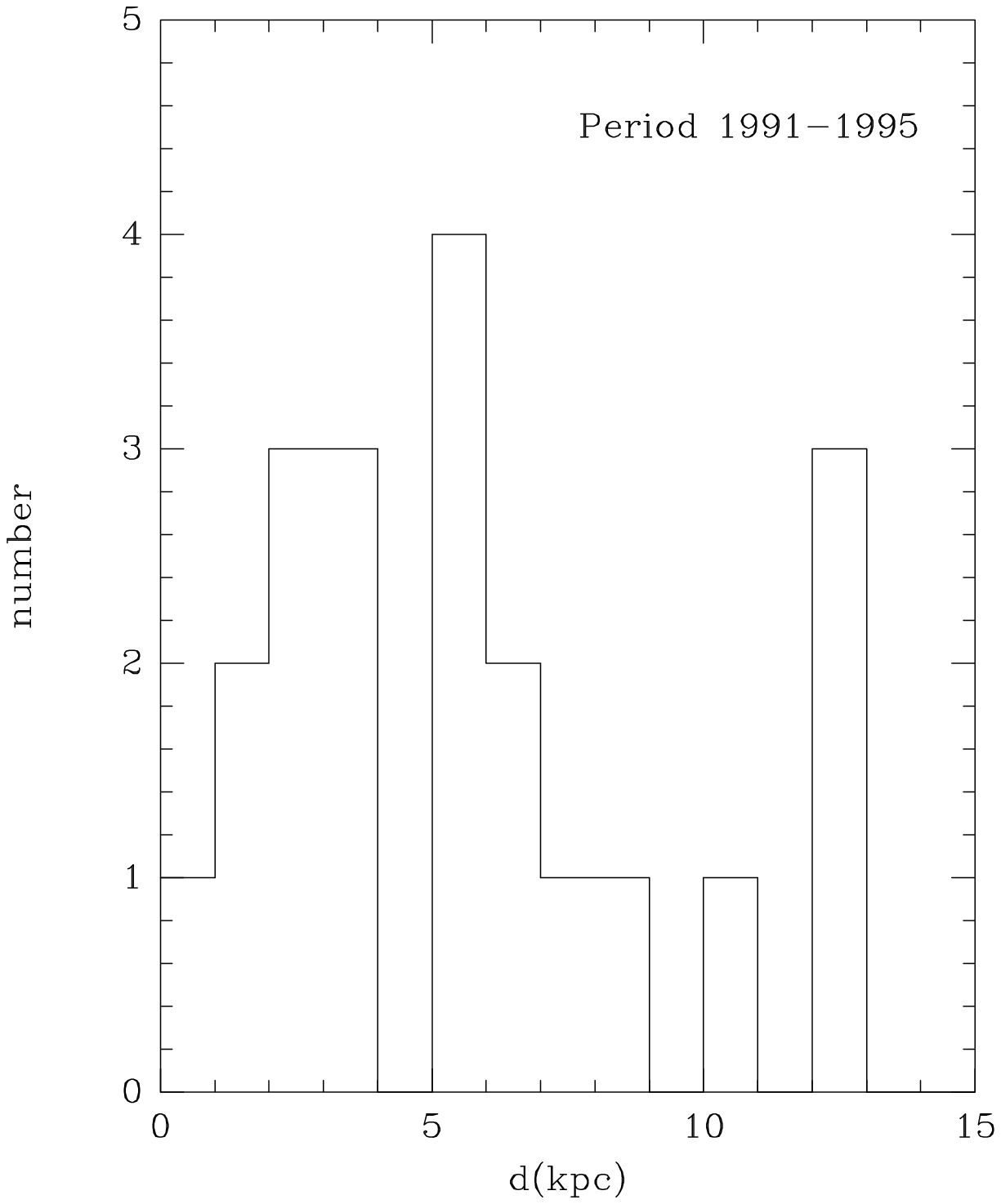}}}
\end{picture}
\caption{(Left) Histogram of novae distances for the novae discovered in the 
last century (until 1995). (Right) Same for the subset of the recent novae 
in the period 1991-1995}
\label{distances}
\end{figure}

In order to estimate the probability of having a nova at a particular 
distance, it is instructive to look at figure \ref{distances}, which shows 
an histogram of the novae distances for the same nova set mentioned above 
\cite{Sha97}, as well as for the subset of novae in the 1991-1995 period. 
The sample of years 1991-1995 suffers from small number statistics, but it is 
more representative of recent more accurate observations. 
Although the distances have a large uncertainty, some 
general trends can be extracted: the observed nova rate for 
novae at distances shorter than 1 kpc is 1/5=0.20 yr$^{-1}$ 
(1991-1995 set), or 16/95=0.17 yr$^{-1}$ 
(complete set 1901-1995), which is not very large. If we relax the distances 
of detectability of novae by INTEGRAL by a factor of 3 (i.e., we adopt 
3 kpc instead of 1 kpc, invoking the effect of the uncertain ejected 
masses -for some observed novae- by a factor of 10), the observed nova rate 
increases to 6/5=1.20 yr$^{-1}$ (1991-1995 nova set), or 50/95=0.53 yr$^{-1}$ 
(complete set 1901-1995). Therefore, there is 
some chance to have a close nova during INTEGRAL's lifetime (2 to 5 years).

Concerning future instrumentation, if an increase of sensitivity by a factor of 
10 (for broad lines) is achieved, the detection of novae would be a routine 
instead of a chance. If, in addition, these instruments have wide fields of view 
and are designed to perform frequent surveys of the sky in the hard X-ray 
domain (E>100 keV), then the prompt e$^-$-e$^+$ annihilation gamma-ray emission 
of novae could be detected for many Galactic novae. This 
fact would be crucial not only for the understanding of the nova explosion 
mechanism itself, but also for the knowledge of the nova distribution in the 
Galaxy (\cite{Har00}. This distribution is not at all known, since only 
3-5 of the 35$\pm$11 Galactic novae exploding every year are discovered optically 
nowadays.   




\begin{table}
\caption{Radioactive isotopes ejected by novae relevant for gamma-ray 
emission}
\begin{tabular}{ccccc}
\hline
\tablehead{1}{c}{}{Isotope}  &  
\tablehead{1}{c}{}{Lifetime} &  
\tablehead{1}{c}{}{Main disintegration process} &
\tablehead{1}{c}{}{Type of $\gamma$-ray emission}  & 
\tablehead{1}{c}{}{Nova type} \\
\hline
$^{13}$N  & 862~s          & $\beta^+$--decay
          & 511~keV line \& continuum      & CO and ONe\\
$^{18}$F  & 158~min        & $\beta^+$--decay
          & 511~keV line \& continuum      & CO and ONe\\
$^{7}$Be  & 77~days        & $e^-$--capture
          & 478~keV line                   & CO\\
$^{22}$Na & 3.75~years     & $\beta^+$--decay
          & 1275~keV \& 511~keV lines      & ONe\\
$^{26}$Al & 10$^{6}$~years & $\beta^+$--decay
          & 1809~keV \& 511~keV lines      & ONe\\
\hline
\end{tabular}
\label{radioac1}
\end{table}


\begin{table}
\caption{Radioactivities in novae ejecta ($^{13}$N and $^{18}$F at 1h after 
T$_{peak}$)}
\begin{tabular}{cccccccc}
\hline
\tablehead{1}{c}{}{Nova} & 
\tablehead{1}{c}{}{M$_{\rm wd}(\rm M_\odot)$}   & 
\tablehead{1}{c}{}{M$_{\rm ejec}(\rm M_\odot)$} &
\tablehead{1}{c}{}{KE (erg/g)}  & 
\tablehead{1}{c}{}{$^{13}$N (M$_\odot$)} & 
\tablehead{1}{c}{}{$^{18}$F (M$_\odot$)} & 
\tablehead{1}{c}{}{$^{7}$Be (M$_\odot$)} & 
\tablehead{1}{c}{}{$^{22}$Na (M$_\odot$)} \\
\hline
CO          & 0.8                       & 6.2x10$^{-5}$                  &
8x10$^{15}$ & 1.5x10$^{-7}$             & 1.8x10$^{-9}$                  &
              6.0x10$^{-11}$            & 7.4x10$^{-11}$                 \\
CO          & 1.15                      & 1.3x10$^{-5}$                  &
4x10$^{16}$ & 2.3x10$^{-8}$             & 2.6x10$^{-9}$                  &
              1.1x10$^{-10}$            & 1.1x10$^{-11}$                 \\
ONe         & 1.15                      & 2.6x10$^{-5}$                  &
3x10$^{16}$ & 2.9x10$^{-8}$             & 5.9x10$^{-9}$                  &
              1.6x10$^{-11}$            & 6.4x10$^{-9}$                  \\
ONe         & 1.25                      & 1.8x10$^{-5}$                  &
4x10$^{16}$ & 3.8x10$^{-8}$             & 4.5x10$^{-9}$                  &
              1.2x10$^{-11}$            & 5.9x10$^{-9}$                  \\
\hline
\end{tabular}
\label{radioac2}
\end{table}


\begin{table}
\caption{SPI 3$\sigma$ detectability distances (in kpc) for lines and 
continuum (see text for details about T$_{obs}$).}
\begin{tabular}{cccccc}
\hline
\tablehead{1}{c}{}{Nova type}     & 
\tablehead{1}{c}{}{M$_{\rm wd}(\rm M_\odot)$} & 
\tablehead{1}{c}{}{511 keV line}  & 
\tablehead{1}{c}{}{478 keV line}  & 
\tablehead{1}{c}{}{1275 keV line} & 
\tablehead{1}{c}{}{(170-470) keV}\\
\hline
CO          & 0.8                       &   0.7        &   0.4        
            & -                         &   0.4\\
CO          & 1.15                      &   2.4        &   0.5        
            & -                         &   2.0\\
ONe         & 1.15                      &   3.7        &   -          
            & 1.1                       &   3.0\\
ONe         & 1.25                      &   4.3        &   -          
            & 1.1                       &   3.0\\
\hline
\end{tabular}
\label{SPIdetect}
\end{table}

\end{document}